\documentclass[pre,aps,twocolumn,showpacs]{revtex4}

\usepackage{dcolumn}
\usepackage{amssymb}
\usepackage{graphicx}

\newcommand \be {\begin{equation}}
\newcommand \ee {\end{equation}}
\newcommand \bea {\begin{eqnarray}}
\newcommand \eea {\end{eqnarray}}
\newcommand \ve {\varepsilon}

\begin{document}

\title{Non-equilibrium temperatures in steady-state systems with conserved energy}

\author{Eric Bertin$^1$, Olivier Dauchot$^2$, and Michel Droz$^1$}
\affiliation{$^1$ Department of Theoretical Physics, University of Geneva, CH-1211 Geneva 4, Switzerland\\
$^2$ SPEC, CEA Saclay, F-91191 Gif-sur-Yvette Cedex, France}
\date{\today}

\begin{abstract}
We study a class of non-equilibrium lattice models describing local
redistributions of a globally conserved quantity, which is interpreted
as an energy.  A particular subclass can be solved exactly, allowing
to define a statistical temperature $T_{th}$ along the same lines as
in the equilibrium microcanonical ensemble. We compute the response
function and find that when the fluctuation-dissipation relation is
linear, the slope $T_{FD}^{-1}$ of this relation differs from the
inverse temperature $T_{th}^{-1}$. We argue that $T_{th}$ is
physically more relevant than $T_{FD}$, since in the steady-state
regime, it takes equal values in two subsystems of a large isolated system.
Finally, a numerical renormalization group procedure suggests that all
models within the class behave similarly at a coarse-grained level,
leading to a new parameter which describes the deviation from
equilibrium.  Quantitative predictions concerning this parameter are
obtained within a mean-field framework.
\end{abstract}

\pacs{05.20.-y, 05.70.Ln, 05.10.Cc}

\maketitle

\section{Introduction}

The existence and the precise definition of intensive thermodynamical
parameters in out-of-equilibrium systems still remains an open
issue.  Indeed, the goal of a statistical approach for
non-equilibrium systems, which remains to be constructed, would be to
give a well-defined meaning to such thermodynamical parameters, and to
predict their relation with extensive macroscopic variables like energy
or volume. Accordingly, many attempts have been made to define
out-of-equilibrium temperatures in the last decades \cite{Jou}. 

In the context of glasses, which are non-stationary systems with very
large relaxation times, effective temperatures have been first
introduced as phenomenological parameters allowing to account for
experimental data \cite{Tool,Naraya,Moynihan}.  More recently, the
notion of effective temperature has been given a more fundamental
status, being defined as the inverse of the slope of
fluctuation-dissipation relations (FDR) in the aging regime
\cite{CuKuPe}.  This definition was guided by the dynamical results
obtained within a family of mean-field spin glass models
\cite{CuKu}. Interestingly, such a definition of the effective
temperature has been shown to satisfy the basic properties expected
for a temperature \cite{CuKuPe}.  Since then, a lot of
numerical simulations
\cite{Parisi,Kob,Barrat00,Sciortino,Berthier,Zamponi,Ritort} and
experiments \cite{Israeloff,Ciliberto,Ocio,Danna,Abou} have
been conducted to
test the validity of this definition of temperature in the aging
regime of glassy materials.  Yet, this definition seems not to be
always applicable as the measured FDR can be non linear.

Other classes of systems are far from equilibrium not due to
their slow relaxation towards the equilibrium state, but rather because
they are subjected to external constraints producing fluxes (of
particles or energy for example) traversing the system,
leading to energy dissipation.
As a result, they never reach an equilibrium state.
Among these systems, one can think of granular gases, sheared fluid
or certain kinetic spin models, to quote only a few of them.

Although the usual formalism of equilibrium statistical physics does not
apply to these systems, it is interesting to note that the latter sometimes
share with equilibrium systems some quantitative properties, like
critical behavior \cite{Grinstein,Tauber}.  To describe the statistical
properties of such non-equilibrium systems, effective temperatures have
been defined either from FDR \cite{Baldassarri,Barrat03,Chate} or from
maximum entropy conditions \cite{Chate,Miller}, as originally proposed by
Jaynes \cite{Jaynes}.  Still, the validity of these procedures remains
to be clarified in the context of non glassy out-of-equilibrium
systems.

When described in a probabilistic language, a common feature
of these systems is that they do
not obey the detailed balance property, considered as a signature of
equilibrium dynamics. 
Since the breaking of detailed balance plays an important role in
non-equilibrium systems, it may be useful to distinguish between
different forms of detailed balance which should not be confused.  In
the literature, the term `detailed balance' often refers to a canonical
form which reads:
\be
W(\beta|\alpha)\, e^{-E_{\alpha}/T} = W(\alpha|\beta)\, e^{-E_{\beta}/T}
\ee
where $W(\beta|\alpha)$ is the transition rate from state $\alpha$ to
state $\beta$.  This criterion on transition rates ensures that the
statistical equilibrium reached at large times by the system is indeed
the canonical equilibrium at temperature $T$.

Still, the above approach requires to know the equilibrium distribution
before defining the stochastic model.  On the contrary, one could try
to find a stochastic model which describes in the best possible way a
given complex hamiltonian system, without knowing a priori the equilibrium
distribution.  Such a stochastic model should at
least preserve the symmetries of the original hamiltonian system, which
are the energy conservation and the time-reversal symmetry 
$t \to -t$ (additional symmetries --translation, rotation, etc.-- must
be taken also into account when present).
Energy conservation is easily implemented in the stochastic rules by
allowing only transitions between states with the same energy.  On the
other side, the time-reversal symmetry in the hamiltonian system can be
interpreted in a stochastic language as the equality between two
opposit transition rates: $W(\beta|\alpha) = W(\alpha|\beta)$, a
property called microcanonical detailed balance or microreversibility.

In the context of non-equilibrium systems, one expects 
that the time-reversal symmetry is broken due to the presence of fluxes
or dissipation.
Hence, a simple way to define a non-equilibrium system
is to consider more general microcanonical forms of detailed balance
relations such as:
\be \label{DBab}
W(\beta|\alpha) f_{\alpha} = W(\alpha|\beta) f_{\beta}
\ee
where $f_{\alpha}$ is the statistical weight of state $\alpha$,
and with $E_{\alpha}=E_{\beta}$.

In this paper, we study a class $\mathcal{C}$ of non-equilibrium
lattice models describing local redistributions of a globally conserved
quantity, which is interpreted as an energy.  A particular subclass
$\mathcal{C}_s$ satisfies a microcanonical detailed balance relation
of the form
(\ref{DBab}), which differs from microreversibility and can be solved
exactly, allowing to define a statistical temperature $T_{th}$ along
the same lines as in the equilibrium microcanonical ensemble.
The response function is computed explicitely, and the FDR is found
to be linear or non linear depending on the model considered --the
response can even be non linear with the perturbing field.
Very interestingly, when the FDR is linear, its slope differs
from the inverse temperature
$T_{th}^{-1}$, which questions the relevance of FDR to define a
temperature in non glassy out-of-equilibrium systems.  Finally, we
implement numerically a functional renormalization group procedure to
argue that all the models within the class $\mathcal{C}$ behave at the
coarse-grained level as a member of the subclass $\mathcal{C}_s$.
Predictions about the renormalization procedure are also made using
mean-field arguments, and are quantitatively verified.  Note that a
short version of some aspects of this work has appeared in \cite{court},
and that a related model, including kinetic constraints, has also
been introduced in the context of glassy dynamics \cite{Lequeux}.

\section{Models and steady-state properties}

\subsection{Definition} \label{Sect-IIdef}

The models we consider in this paper are defined as follows.  On each
site $i$ of a $d$-dimensional lattice, a real variable $x_i$ which can
take either positive or negative values is introduced.  Dynamical rules
are defined such that the quantity
\be
E=\sum_{i=1}^N g(x_i)
\ee
is conserved. The function $g(x)$, assumed to be positive with continuous
derivative, decreases for $x<x_0$ and
increases for $x>x_0$, where $x_0$ is an arbitrary given value.
Without loss of generality, we assume $g(x_0)=0$.
Quite importantly, the steady-state distribution can be
computed with these hypotheses only.
However, to clarify the presentation, we
assume in this section that $x_0=0$ and $g(x)$ is an even function of $x$.

It is also necessary to introduce the reciprocal function $g^{-1}(y)$,
given as the positive root of the equation $g(x)=y$.  The dynamics is
defined as follows: at each time step, a link $(j,k)$ is randomly
chosen on the lattice and the corresponding variables $x_j$ and $x_k$
are updated so as to conserve the energy $g(x_j)+g(x_k)$ of the link.
To be more specific, the new values $x_j'$ and $x_k'$ are given by
\be \label{def-model}
x_j' = \pm g^{-1} (q S_{jk}) \qquad x_k' = \pm g^{-1} ((1-q) S_{jk})
\ee
with $S_{jk} \equiv g(x_j)+g(x_k)$, and $q$ is a random variable drawn
from a distribution $\psi(q)$, assumed to be symmetric with respect to
$q=\frac{1}{2}$ ($0<q<1$).  The new values $x_j'$ and $x_k'$ are either
positive or negative with equal probability, and without correlation
between the signs.  Thus a model belonging to this class is
characterized by two functions $g(x)$ and $\psi(q)$.

\subsection{Master equation}

The system is described by the distribution $P(\{x_i\},t)$, which gives
the probability to be in a configuration $\{x_i\}$ at time $t$.  Its
evolution is given by the master equation:
\bea \label{eqMEdef}
\frac{\partial P}{\partial t}(\{x_i\},t) &=& \int \prod_i dx_i' \,
W(\{x_i\}|\{x_i'\}) \, P(\{x_i'\},t) \\ \nonumber
&& \quad - \int \prod_i dx_i' \, W(\{x_i'\}|\{x_i\}) \, P(\{x_i\},t)
\eea
where $W(\{x_i'\}|\{x_i\})$ is the transition rate from configuration
$\{x_i\}$ to configuration $\{x_i'\}$.  The transition rate can be
decomposed into a sum over the links of the lattice:
\be
W(\{x_i'\}|\{x_i\}) = \sum_{\langle j,k \rangle} W_{jk}(\{x_i'\}|\{x_i\})
\ee
where $W_{jk}(\{x_i'\}|\{x_i\})$ accounts for the redistribution over
a given link $(j,k)$:
\bea \label{def-W}
&&W_{jk}(\{x_i'\}|\{x_i\}) = \left[ \prod_{i \ne j,k} \delta(x_i'-x_i)
\right] \int_0^1 dq \, \psi(q) \times \\ \nonumber
&& \: \frac{1}{4} \sum_{\sigma_j, \sigma_k} \delta \left[ x_j' - \sigma_j
g^{-1}(qS_{jk}) \right] \, \delta \left[ x_k' - \sigma_k g^{-1}((1-q)S_{jk})
\right]
\eea
where the variables $\sigma_j$, $\sigma_k = \pm 1$ account for the
random signs appearing in Eq.~(\ref{def-model}) with probability $\frac{1}{2}$
--hence the factor $\frac{1}{4}$ in the above equation.
After some algebra, the transition rate $W(\{x_i'\}|\{x_i\})$
can be rewritten as:
\bea
&&W(\{x_i'\}|\{x_i\}) = \frac{1}{4} \sum_{\langle j,k \rangle}
\left[ \prod_{i \ne j,k} \delta(x_i'-x_i) \right] \,\times \\ \nonumber
&& \quad \frac{|g'(x_j') g'(x_k')|}{S_{jk}} \, \psi
\left( \frac{g(x_j')}{S_{jk}} \right) \delta \left[ g(x_j') + g(x_k')
- S_{jk} \right]
\eea
where $g'(x)$ denotes the derivative of $g(x)$.

\subsection{Detailed balance and steady-state distribution}
\label{sect-steady}

A case of particular interest is the subclass of models for which the distribution $\psi(q)$ is given by a symmetric beta law:
\be
\psi(q) = \frac{\Gamma(2\eta)}{\Gamma(\eta)^2} \, q^{\eta-1} (1-q)^{\eta-1}
\ee
with $\eta>0$.
In this case, the function $\psi(g(x_j')/S_{jk})$ appearing in the
transition rates factorizes, if one takes into account the delta
function.  So the transition rate reads
\bea
&& W(\{x_i'\}|\{x_i\}) = \frac{\Gamma(2\eta)}{4 \Gamma(\eta)^2} \;
\sum_{\langle j,k \rangle} \left[ \prod_{i \ne j,k} \delta(x_i'-x_i) \right] \times \\ \nonumber
&& \frac{|g'(x_j') g'(x_k')|}{S_{jk}^{2\eta-1}}
\, g(x_j')^{\eta-1} g(x_k')^{\eta-1} \delta \left[ g(x_j') + g(x_k') - S_{jk} \right]
\eea
From this last expression, it can be checked that a detailed balance relation is satisfied:
\bea
&& W(\{x_i'\}|\{x_i\}) \, \prod_{i=1}^N \left[ |g'(x_i)| \, g(x_i)^{\eta-1} \right] = \\ \nonumber
&& \qquad \qquad W(\{x_i\}|\{x_i'\}) \, \prod_{i=1}^N \left[ |g'(x_i')| \, g(x_i')^{\eta-1} \right]
\label{eq-DB}
\eea
As a result, the steady-state distribution $P_{st}(\{x_i\}|E)$, for a given value $E$ of the energy, is readily obtained as:
\bea
\label{PxkE}
P_{st}(\{x_i\}|E) &=& \frac{1}{Z_N(E)} \, \prod_{i=1}^N \left[ |g'(x_i)| \, g(x_i)^{\eta-1} \right] \times \\ \nonumber
&& \qquad \qquad \qquad \quad \delta \left( \sum_{i=1}^N g(x_i) - E \right)
\eea
where $Z_N(E)$ is a normalization factor that may be called an effective (microcanonical) partition function:
\bea
\label{eq-ZNE}
Z_N(E) &=& \int \prod_{i=1}^N \left[ dx_i \, |g'(x_i)| \, g(x_i)^{\eta-1} \right] \times \\ \nonumber
&& \qquad \qquad \qquad \quad \delta
\left( \sum_{i=1}^N g(x_i) - E \right)
\eea
An important remark has to be made at this stage: Eqs.~(\ref{PxkE}) and
(\ref{eq-ZNE}) remain formally valid if one slightly changes the
definition of the model.  This can be done in two different ways.
First, one could consider the case where the variables $\{x_i\}$ take only
positive values.  Then one only needs to remove the sum
$\frac{1}{4}\sum_{\sigma_j,\sigma_k}$ in the transition rates given in
Eq.~(\ref{def-W}), and Eq.~(\ref{PxkE}) is recovered, with this time
$x_i>0$.  Second, as mentioned in Sect.~\ref{Sect-IIdef},
the model can be generalized by assuming that $g(x)$
is not an even function.  This is particularly useful if one wants to
include an external field which breaks the $+/-$ symmetry --see
Sect.~\ref{sect-FDR}.  Actually, if $g(x)$ decreases for $x<x_0$,
and increases for $x>x_0$, the
distribution given in Eq.~(\ref{PxkE}) also holds
\footnote{In this case, it is necessary to introduce two different
reciprocal functions, $g_-^{-1}(y)$ which takes values in
$(-\infty,x_0]$, and $g_+^{-1}(y)$ which takes values in
$[x_0,+\infty)$.  In the redistribution process, each of these two
intervals is chosen with equal probability.}.

The function $Z_N(E)$ can be computed using a Laplace transform.
Indeed, it appears rather clearly from Eq.~(\ref{eq-ZNE}), by making
the change of variable $\ve_i = g(x_i)$, that $Z_N(E)$ is actually
independent of the functional form of $g(x)$. One finds
\be \label{eqZNE}
Z_N(E) = \kappa_N \, E^{\eta N - 1}
\ee
with $\kappa_N = 2^N \, \Gamma(\eta)^N / \Gamma(\eta N)$.
The fact that $Z_N(E)$ does not depend on $g(x)$ is actually not a
coincidence, but comes from the basic definition of the model given in
Eq.~(\ref{def-model}). Indeed, for any function $g(x)$, one could choose
as the dynamical variables the local energies $\ve_i = g(x_i)$, and
solve the model for $\ve_i$. Coming back to the variable $g(x_i)$ at the
end of the calculations, the distribution (\ref{PxkE}) would be recovered.
Still, it should not be concluded from this that all physical quantities
defined in the model are independent of $g(x)$. In particular, the
response to a perturbing field depends strongly on $g(x)$, since the
field is coupled to $x$, and not to the energy $g(x)$ --see
Sect.~\ref{sect-FDR}.

An interesting question is also to see under what conditions
microreversibility (to be associated to the equilibrium behavior) can
be recovered in this model.  Microreversibility holds if $|g'(x)| \,
g(x)^{\eta-1}$ is independent of $x$, as can be seen from
Eq.~(\ref{eq-DB}).  Such a condition can be satisfied only if $g(x)$ is
a power law, say $g(x) = x^p/p$, where $p$ is an even integer to ensure the
regularity of $g(x)$ around $x=0$.  The factor $1/p$ has been added for
convenience, but is otherwise arbitrary.
One then has
\be
|g'(x)| \, g(x)^{\eta-1} = p^{1-\eta} \, |x|^{\eta p-1}
\ee
Accordingly, microreversibility is recovered for $\eta=1/p$. On the
contrary, for $\eta \ne 1/p$, significant differences with the
equilibrium behavior are expected.  These differences may be even
stronger if $g(x)$ is not a power law.

\section{Non-equilibrium temperatures}

\subsection{Statistical approach}

\subsubsection{Microcanonical equilibrium}

In order to define a temperature in this model, one can try to follow a
procedure similar to that of the microcanonical ensemble in equilibrium
statistical physics.  Indeed, one of the main motivations when building
the present model was to find a model in which a global quantity (the
energy) is conserved, so as to `mimic' in some sense a microcanonical
situation.  Yet, as mentioned above, the absence of microreversibility
should yield important differences with the latter
case.  For an equilibrium system in the microcanonical ensemble,
temperature is introduced in the following way.  Considering a large
system $\mathcal{S}$ with fixed energy, one introduces a partition
into two subsystems $\mathcal{S}_1$ and $\mathcal{S}_2$, with energy
$E_\ell$ and a number $N_\ell$ of degrees of freedom ($\ell=1,2$).
These two subsystems are no longer isolated, since they can mutually
exchange energy; the only constraint is that $E_1+E_2=E$ is fixed.  The
key quantity is then the number $\Omega_{N_\ell}(E_\ell)$ of accessible
states with energy $E_\ell$ in the subsystem $\mathcal{S}_\ell$; in
systems with continuous degrees of freedom (like a classical gas for
instance),  $\Omega_{N_\ell}(E_\ell)$ is the area of the hypersurface
of energy $E_\ell$ in phase space.  Assuming that both subsystems do
not interact except by exchanging energy, the number of states of the
system $\mathcal{S}$ compatible with the partition $(E_1,E_2)$ of the
energy is equal to $\Omega_{N_1}(E_1) \Omega_{N_2}(E_2)$.  But since
$E_1+E_2$ is fixed, the most probable value $E_1^*$ is found from the
maximum, with respect to $E_1$, of $\Omega_{N_1}(E_1)
\Omega_{N_2}(E-E_1)$.  Taking a logarithmic derivative, one finds the
usual result:
\be \label{eqTmic}
\frac{\partial \ln \Omega_{N_1}}{\partial E_1} \Big\vert_{E_1^*} = \frac{\partial \ln \Omega_{N_2}}{\partial E_2} \Big\vert_{E-E_1^*}
\ee
Defining the microcanonical temperature $T_\ell$ of subsystem $\ell$ by the relation
\be
\frac{1}{T_{\ell}} = \frac{\partial \ln \Omega_{N_\ell}}{\partial E_\ell} \Big\vert_{E_\ell^*}
\ee
one sees from Eq.~(\ref{eqTmic}) that $T_1=T_2$, i.e. that the
temperatures are equal in both subsystems (throughout the paper, the
Boltzmann constant $k_B$ is set to unity).  In addition, it can also be
shown that the common value $T$ does not depend on the partition
chosen; as a result, $T$ is said to characterize the full system
$\mathcal{S}$.

\subsubsection{`Microcanonical' stationary state}

Very interestingly, this microcanonical definition of temperature can
be generalized in a rather straightforward way to the present model.
Still, it should be noticed first that microscopic configurations
compatible with the given value of the energy are no longer
equiprobable, as seen from the distribution (\ref{PxkE}), so that
$\Omega_N(E)$ is no more relevant to the problem.  But starting again
from a partition into two subsystems as above, one can determine the
most probable value $E_1^*$ from the maximum of the conditional
probability $P(E_1|E)$ that subsystem $\mathcal{S}_1$ has energy $E_1$
given that the total energy is $E$.  Indeed, in the equilibrium case,
$P(E_1|E)$ reads
\be \label{PE1Emic}
P(E_1|E) = \frac{\Omega_{N_1}(E_1) \, \Omega_{N_2}(E-E_1)}{\Omega_N(E)}
\ee
which by derivation with respect to $E_1$, yields precisely the same result as Eq.~(\ref{eqTmic}).

To be more specific, the subsystems are defined in the present model as
a partition of the lattice, with $N_1$ sites in $\mathcal{S}_1$ and
$N_2$ sites in $\mathcal{S}_2$.  The conditional distribution
$P(E_1|E)$ is then given by:
\bea \nonumber
P(E_1|E) &=& \int \prod_{i=1}^N dx_i \, P_{st}(\{x_i\}|E) \, \delta \left( \sum_{i \in \mathcal{S}_1} g(x_i) - E_1 \right) \\
&=& \frac{1}{Z_N(E)} \int \prod_{i=1}^N \left[ dx_i \, |g'(x_i)| \, g(x_i)^{\eta-1} \right] \times \\ \nonumber
&& \quad \delta \left( \sum_{i=1}^N g(x_i) - E \right) \, \delta \left( \sum_{i \in \mathcal{S}_1} g(x_i) - E_1 \right)
\eea
Taking into account the last delta function, the first one can be replaced by $\delta ( \sum_{i \in \mathcal{S}_2} g(x_i) - (E-E_1) )$,
so that $P(E_1|E)$ may be written in a compact form as:
\be \label{eq-PE1E}
P(E_1|E) = \frac{Z_{N_1}(E_1) \, Z_{N_2}(E-E_1)}{Z_N(E)}
\ee
This result generalizes in a nice way the equilibrium distribution Eq.~(\ref{PE1Emic}), since in equilibrium $Z_N(E)$ reduces precisely to $\Omega_N(E)$.
The most probable value $E_1^*$ satisfies
\be
\frac{\partial \ln P(E_1|E)}{\partial E_1} \Big\vert_{E_1^*} = 0
\ee
which yields
\be \label{eqTmodel}
\frac{\partial \ln Z_{N_1}}{\partial E_1} \Big\vert_{E_1^*} = \frac{\partial \ln Z_{N_2}}{\partial E_2} \Big\vert_{E-E_1^*}
\ee
So in close analogy with the equilibrium approach, we define a temperature $T_{th}^\ell$ for subsystem $\mathcal{S}_{\ell}$ through
\be \label{def-Tth-ell}
\frac{1}{T_{th}^\ell} = \frac{\partial \ln Z_{N_\ell}}{\partial E_\ell} \Big\vert_{E_\ell^*}
\ee
Then Eq.~(\ref{eqTmodel}) implies that $T_{th}^1 = T_{th}^2$.

At this stage, it is important to check that the common value $T_{th}$
of the temperature does not depend on the partition chosen.  To this
aim, we show that $T_{th}$ can be expressed as a function of global
quantities characterizing the whole system, with no reference to the
specific partition.

Let us compute $Z_N(E)$ as a function of $Z_{N_1}(E_1)$ and $Z_{N_2}(E_2)$.
Since $\int_0^E dE_1 \, P(E_1|E) = 1$, one has from Eq.~(\ref{eq-PE1E}):
\be
Z_N(E) = \int_0^E dE_1 \, Z_{N_1}(E_1) \, Z_{N_2}(E-E_1)
\ee
We assume the following general scaling form at large $N_{\ell}$ for $Z_{N_{\ell}}(E_{\ell})$:
\be \label{scal-ZNE}
Z_{N_{\ell}}(E_{\ell}) = A_{\ell} \, \exp[N_{\ell} \zeta_{\ell}(\overline{\ve}_{\ell})]
\ee
with $\overline{\ve}_{\ell} \equiv E_{\ell}/N_{\ell}$ (the index $\ell=1,2$ labels the subsystem).
This scaling form is demonstrated explicitely in Sect.~\ref{sect-entropy}.
Using a saddle-point calculation, one obtains for $Z_N(E)$ a relation of the form (with $\overline{\ve} = E/N$):
\be
Z_N(E) = Z_{N_1}(E_1^*) \, Z_{N_2}(E_2^*) \, N \int_0^{\overline{\ve}} d\ve_1 \, e^{-N \, b(\overline{\ve}) \, (\ve_1 - \overline{\ve}_1^*)^2}
\ee
where $b(\overline{\ve})$ is defined as:
\be
b(\overline{\ve}) = -\frac{1}{2} \left[ \lambda_1 \zeta_1''(\overline{\ve}_1^*) + \lambda_2 \zeta_2''(\overline{\ve}-\overline{\ve}_1^*) \right]
\ee
with $\lambda_{\ell} = N_{\ell}/N$.
Thus $\ln Z_N(E)$ reads:
\be
\ln Z_N(E) = \ln Z_{N_1}(E_1^*) +  \ln Z_{N_2}(E-E_1^*) - \frac{1}{2} \ln b(\overline{\ve}) + C
\ee
where $C$ does not depend on $E$.
Taking the derivative with respect to $E$ yields, using Eq.~(\ref{def-Tth-ell}):
\be
\frac{\partial \ln Z_N}{\partial E}
= \frac{1}{T_{th}} \, \frac{\partial E_1^*}{\partial E} + \frac{1}{T_{th}}
\, \left( 1 - \frac{\partial E_1^*}{\partial E} \right) - \frac{1}{2N} \, b'(\overline{\ve})
\ee
In the limit $N \to \infty$ (with $\overline{\ve}$ fixed), the last term
vanishes, whereas $T_{th}$ has a finite limit due to the scaling form
Eq.~(\ref{scal-ZNE}), so that
\be \label{eq-defTth}
\frac{\partial \ln Z_N}{\partial E} = \frac{1}{T_{th}}
\ee
As a result, $T_{th}$ can be computed from the global quantity $Z_N(E)$
instead of $Z_{N_1}(E_1)$ or $Z_{N_2}(E_2)$, and is thus independent
of the partition chosen.  This temperature characterizes the
statistical state of the whole system.  From Eq.~(\ref{eqZNE}), the
equation of state of the system is:
\be \label{eq-state}
E = \eta N T_{th}
\ee
In the case of a quadratic energy, i.e.~$g(x)=\frac{1}{2} x^2$, it has
been shown above that the equilibrium behavior is recovered for
$\eta=\frac{1}{2}$.  This result is confirmed by Eq.~(\ref{eq-state}),
which reduces for $\eta=\frac{1}{2}$ to the usual form of the energy
equipartition.  On the contrary, for $\eta \ne \frac{1}{2}$, a
generalized form of equipartition holds in the sense that all the sites
have the same average energy $\overline{\ve} = E/N$ (which is not
surprising given the homogeneity of the system), but this average
energy per degree of freedom is equal to $\eta T_{th}$ instead of
$\frac{1}{2} T_{th}$.  This point will be discussed in more details
later on.

Up to now, we have considered only the `microcanonical'  (in a
generalized sense) distribution $P_{st}(\{x_i\}|E)$.  Yet, it would be
interesting to introduce also the analogous of the canonical
distribution. To do so, we compute the distribution $P_{can}(\{x_i\})$
associated to a small (but still macroscopic) subsystem $\mathcal{S}_1$
of a large isolated system $\mathcal{S}$.
The degrees of freedom $\{x_i\}$ with $i=N_1+1, \ldots, N$ have to be
integrated out since they belong to the reservoir. One finds for the
remaining $\{x_i\}$ ($i=1, \ldots, N_1$) the following distribution:
\bea \label{eq-Pcan0}
&& P_{can}(\{x_i\}) = \frac{1}{Z_N(E)} \prod_{i=1}^{N_1} |g'(x_i)| \
 g(x_i)^{\eta-1} \times \\ \nonumber
&& \int \prod_{i=N_1+1}^N \left[ dx_i \, |g'(x_i)|
\, g(x_i)^{\eta-1} \right] \, \delta \left( \sum_{i=1}^N g(x_i) - E \right)
\eea
The above integral is nothing but the partition function $Z_{N_2}(E-\sum_{i=1}^{N_1} g(x_i))$, with $N_2=N-N_1$,
which can be expanded to first order as:
\be
\ln Z_{N_2} \left(E-\sum_{i=1}^{N_1} g(x_i)\right) =
\ln Z_{N_2}(E) - \frac{1}{T_{th}} \sum_{i=1}^{N_1} g(x_i)
\ee
assuming that $\sum_{i=1}^{N_1} g(x_i) \ll E$, which is true as long as
$N_1 \ll N$. The derivative of $\ln Z_{N_2}(E)$ has been identified with
$1/T_{th}$ using Eq.~(\ref{def-Tth-ell}), up to corrections that vanish in the limit $N_1/N \to 0$, since $E$ is the total energy rather than the energy
$E_2$ of the reservoir.
Introducing this last result into Eq.~(\ref{eq-Pcan0}), one finally finds
\bea \label{eq-Pcan}
P_{can}(\{x_i\}) &=& \frac{1}{Z_{N_1}^{can}} \, \prod_{i=1}^{N_1} |g'(x_i)| \, g(x_i)^{\eta-1} \times \\  \nonumber
&& \qquad \qquad  \exp \left( - \frac{1}{T_{th}} \sum_{i=1}^{N_1}
g(x_i) \right)
\eea
where $Z_{N_1}^{can} = Z_{N_2}(E)/Z_N(E)$ --note that $E$ is the energy of the
global system which includes the reservoir.
This `canonical' distribution appears to be useful in order to compute 
the FDR, as discussed below in Sect.~\ref{sect-FDR}.

\subsubsection{Entropy and thermodynamics}
\label{sect-entropy}

From Eq.~(\ref{eq-defTth}), it is tempting to generalize the notion of microcanonical entropy through $S(E)=\ln Z_N(E)$.
Indeed, this definition is not only an analogy, but as we shall see, it can be associated with a time-dependent entropy which is maximized by the dynamics.
To define the entropy, one needs first to introduce the probability measure $P_E(\{x_i\},t)$ restricted to the hypersurface of energy $E$:
\be
P(\{x_i\},t) = P_E(\{x_i\},t) \; \delta \left( \sum_{i=1}^N g(x_i)-E \right)
\ee
Then the dynamical entropy is defined as:
\be \label{SEt-def}
S_E(t) = - \int \prod_{i=1}^N dx_i \, P(\{x_i\},t) \, \ln \frac{P_E(\{x_i\},t)}{f(\{x_i\})}
\ee
where $f(\{x_i\}) \equiv \prod_{i=1}^N |g'(x_i)| \, g(x_i)^{\eta-1}$.
Using the master equation (\ref{eqMEdef}), it can be shown that
$S_E(t)$ is a non-decreasing function of time --see
Appendix~\ref{ap-entropy}. As a result, $S_E(t)$ is maximal in the
stationary state, and the corresponding value $S(E)$ is given by:
\bea \nonumber
S(E) &=& - \int \prod_{i=1}^N dx_i \, P(\{x_i\}|E) \, \ln \frac{1}{Z_N(E)} \\
&=& \ln Z_N(E)
\eea
which matches exactly the definition proposed above on the basis of Eq.~(\ref{eq-defTth}).

Using Eq.~(\ref{eqZNE}), one can compute $S(E)$ and check explicitely
that the entropy per site $S(E)/N$ becomes in the thermodynamic limit a
well-defined function $\zeta(\overline{\ve})$ of the energy density $\overline{\ve}=E/N$.
The entropy $S(E)$ reads:
\be
S(E) = N \left[ \ln 2\Gamma(\eta) - \frac{1}{N} \ln \Gamma(\eta N) + \eta \ln E \right]
\ee
As $\ln \Gamma(x) \approx x\ln x-x$ for large $x$, one finds:
\be
\frac{1}{N} \ln \Gamma(\eta N) \approx \eta (\ln \eta -1) + \eta \ln N
\ee
which allows to write $S(E)=N\zeta(\overline{\ve})$ with:
\be \label{sigma-eps}
\zeta(\overline{\ve}) = \eta \ln \overline{\ve} + \ln 2\Gamma(\eta) - \eta (\ln \eta -1)
\ee

On the other hand, the equilibrium thermodynamic formalism is most
often formulated in terms of the canonical ensemble.  In the present
model, since a canonical distribution has been derived,
it may also be possible to define an equivalent
of the canonical thermodynamic formalism.  Indeed, from
Eq.~(\ref{eq-Pcan}), one can easily see that the average
energy $\langle E \rangle$ is given by
\be
\langle E \rangle = - \frac{\partial \ln Z_N^{can}}{\partial \beta}
\ee
where $\beta \equiv T_{th}^{-1}$ is the inverse temperature.
A generalized free energy $F(T_{th})$ is also naturally introduced through
\be
F(T_{th}) = - T_{th} \ln Z_N^{can}
\ee
The generalized partition function $Z_N^{can}$ can be easily computed, as it is factorized:
\bea \nonumber
Z_N^{can} &=& \left[ \int_{-\infty}^{\infty} dx \, |g'(x)| \, g(x)^{\eta-1} e^{-g(x)/T_{th}} \right]^N \\
&=& \left[ 2 \int_0^{\infty} d\ve \, \ve^{\eta-1} \, e^{-\ve/T_{th}} \right]^N
\eea
which leads to
\be
Z_N^{can} = \left( 2\Gamma(\eta) T_{th}^{\eta} \right)^N
\ee
So the free-energy is given by
\be
F = -NT_{th} \left[ \ln 2\Gamma(\eta) + \eta \ln T_{th} \right]
\ee
In equilibrium, the entropy $S$ is related to the free energy $F$ through
\be
\frac{\partial F}{\partial T} = - S
\ee
This relation is also satisfied within the present model:
\bea \nonumber
\frac{\partial F}{\partial T_{th}} &=& -N \left[ \eta (\ln T_{th} + 1) + \ln 2\Gamma(\eta) \right] \\
&=& -N \zeta(\overline{\ve})
\eea
where the last equality is obtained by using the equation of state $T_{th} = \overline{\ve}/\eta$, and comparing with Eq.~(\ref{sigma-eps}).

\subsection{Fluctuation-dissipation relations}
\label{sect-FDR}

As recalled in the introduction, temperatures are usually defined in
out-of-equilibrium systems as the inverse slope of the FDR, when this
relation is linear.  This approach has been shown to be physically
meaningful in the context of glassy models in the aging regime \cite{CuKu}.
In this case, the long time slope of the FDR gives an effective
temperature which differs from the heat bath temperature.  Still, for
non-equilibrium steady-state systems which are not glassy, no
justification has been proposed to show that the inverse slope of the
FDR satisfies the basic properties expected for a temperature. For
instance, one expects a temperature to take equal values in two
subsystems of a large system, when the stationary state has been
reached.  The present model thus allows to test explicitely the validity
of the FDR definition of temperature.

A natural observable to consider in this model is
\be
M(t) = \sum_{i=1}^N x_i(t)
\ee
The steady-state correlation function $C(t)$ of the system is then
defined as the normalized autocorrelation of the observable $M(t)$
between time $t=0$ and $t$:
\be
C(t) = \frac{1}{N} \langle ( M(t) - \langle M \rangle ) \, ( M(0) - \langle M \rangle ) \rangle
\ee
where the brackets $\langle \dots \rangle$ denote an average over all
possible trajectories of the system.  Calculations are easier using the
canonical distribution $P_{can}(\{x_i\})$; since this distribution
is factorized, the random variables $x_i$ and $x_j$
are independent if $i \ne j$, so that $C(t)$ reduces to
\be \label{eq-cort}
C(t) = \langle ( x(t) - \langle x \rangle ) \, ( x(0) - \langle x \rangle ) \rangle
\ee
where $x$ stands for any of the variables $x_i$ --all sites have the same average values.

The aim of the FDR is to relate correlation and response of a given
observable.  One thus needs to introduce a perturbation which generates
variations of $x$ so that a response could be defined.  A simple way to
perturb the system is to add to the energy a linear term proportional
to an external field $h$: one then replaces $E$ by $E_h$ defined as:
\be
E_h = \sum_{i=1}^N g_h(x_i) = \sum_{i=1}^N g(x_i) - h x_i + c_h
\ee
Without loss of generality, the new function $g_h(x)$ is shifted by a
constant $c_h$ so that the minimum value of $g_h(x)$ remains equal to
$0$.  If the second derivative $g''(0)$ does not vanish,
$c_h$ is given to leading order in $h$ by:
\be
c_h = \frac{h^2}{2 \, g''(0)}
\ee
In order to define the response function, one assumes that the system
is subjected to a field $h \ne 0$ for $t<0$, and that it has reached a
steady state.  Then at time $t=0$, the field $h$ is switched off.  The
(time-dependent) response is defined for $t>0$ through:
\be
\chi(t) \equiv \frac{\partial}{\partial h} \Big\vert_{h=0} \left< \frac{1}{N} \sum_{i=1}^N x_i(t) \right>_h
\ee
where the index $h$ on the brackets indicate that the average is taken
over the dynamics in presence of the field $h$.  The observable
$\langle N^{-1} \sum_i x_i(t) \rangle_h$ can be computed as:
\bea
\left< \frac{1}{N} \sum_{i=1}^N x_i(t) \right>_h &=& \int \prod_{i=1}^N dx_i \, dx_i' \, G_t^0(\{x_i\}|\{x_i'\}) \, \times \\ \nonumber
&& \qquad \qquad P_{can}(\{x_i'\},h) \left( \frac{1}{N} \sum_{i=1}^N x_i \right)
\eea
where $G_t^0(\{x_i\}|\{x_i'\})$ is the zero-field Green function,
i.e.~the probability for the system to be in a configuration $\{x_i\}$
at time $t$, given that it was in a configuration $\{x_i'\}$ at time
$t=0$, in the absence of field.  The response function $\chi(t)$ is
obtained by taking the derivative of the above equation with respect to
$h$, at $h=0$:
\bea \label{chit1}
\chi(t) &=& \int \prod_{i=1}^N dx_i \, dx_i' \, G_t^0(\{x_i\}|\{x_i'\}) \, \times \\ \nonumber
&& \qquad P_{can}(\{x_i'\},0) \, \frac{\partial \ln P_{can}}{\partial h}(\{x_i'\},0) \left( \frac{1}{N} \sum_{i=1}^N x_i \right)
\eea
The canonical distribution $P_{can}(\{x_i\},h)$ in the presence of field takes the same form as Eq.~(\ref{eq-Pcan}), simply replacing $g(x)$ by $g_h(x)$.
Thus one finds for the logarithmic derivative of $P_{can}(\{x_i\},h)$:
\bea \label{dlnPdh}
\frac{\partial \ln P_{can}}{\partial h}(\{x_i'\},0) &=& - \frac{\partial \ln Z_N^{can}}{\partial h} \Big\vert_{h=0} \, + \\ \nonumber
&& \sum_{i=1}^N \left( \frac{x_i}{T_{th}} - \frac{1}{g'(x_i)} - (\eta-1) \frac{x_i}{g(x_i)} \right)
\eea
Note that $dc_h/dh = 0$ at $h=0$, due to the regularity of $g(x)$.
The derivative of the partition function yields:
\be \label{dlnZdh}
\frac{\partial \ln Z_N^{can}}{\partial h} \Big\vert_{h=0} = \frac{1}{T_{th}} \left< \sum_{i=1}^N x_i \right> - \left< \sum_{i=1}^N \omega_i \right>
\ee
where $\omega_i$ stands for
\be
\omega_i \equiv \frac{1}{g'(x_i)} + (\eta-1) \frac{x_i}{g(x_i)}
\ee
Replacing the expression (\ref{dlnPdh}) in Eq.~(\ref{chit1}), one finally finds, using the factorization of the canonical distribution:
\be
\chi(t) = \frac{1}{T_{th}} \, C(t) - \langle (x(t) - \langle x \rangle) (\omega(0) - \langle \omega \rangle) \rangle
\ee
where indices are omitted just as in Eq.~(\ref{eq-cort}).
Compared to the usual form of FDR, an additional term appears which corresponds to the correlation of the variables $x$ and $\omega$.
In general, this new correlation function is not proportional to $C(t)$,
so that a parametric plot of $\chi(t)$ versus $C(t)$, usually referred to as a
fluctuation-dissipation plot, would be non linear.

Yet, in the case where $g(x)$ is an even function of $x$, some important simplifications occur.
On the one hand, the average values of $x$ and $\omega$ vanish.
On the other hand, the correlation $x(t) \omega(0)$ becomes proportional to the `hopping correlation function' $\Phi(t)$, defined as
\be
\Phi(t) = \left< \frac{1}{N} \, \sum_{i=1}^N \phi_i(t) \right>
\ee
The variables $\phi_i(t)$ are history dependent random variables, which
are equal to $1$ if no redistribution involving site $i$ occured
between $t=0$ and $t$, and are equal to $0$ otherwise.  The
proportionality of both correlation functions can be understood as
follows:  if there was a redistribution on site $i$ between $0$ and
$t$, $x_i(t)$ becomes fully decorrelated from $\omega_i$, due to the
fact that the sign of $x_i(t)$ is chosen at random, and that the average
values $\langle x \rangle$ and $\langle \omega \rangle$ vanish for an
even $g(x)$.
On the contrary, if no redistribution occured, $x_i(t) \omega_i(0) =
x_i(0) \omega_i(0)$.  The same reasoning also holds for $C(t)$, so that
one has:
\be
C(t) = \langle x^2 \rangle \, \Phi(t), \quad
\langle x(t) \omega(0) \rangle = \langle x \omega \rangle \, \Phi(t)
\ee
As a result, the FDR can be expressed, in the case of an even function $g(x)$, as
\be
\chi(t) = \left( \frac{1}{T_{th}} - \frac{\langle x \omega \rangle}{\langle x^2 \rangle} \right) C(t)
\ee
So the FDR is indeed linear in this case, and one can define an effective temperature $T_{FD}$ from the inverse slope of this relation.
This yields:
\be \label{defTFD}
\frac{1}{T_{FD}} = \frac{1}{T_{th}} - \frac{\langle x \omega \rangle}{\langle x^2 \rangle}
\ee
Still, as long as $\langle x \omega \rangle \ne 0$, the temperature
$T_{FD}$ differs from the temperature $T_{th}$ defined above from
statistical considerations --a more detailed discussion on this point is
given below in Sect.~\ref{sect-relevance}.

Even though the two temperatures are not equal, one can wonder whether
they are proportional, in the sense that the ratio $T_{th}/T_{FD}$ would be
independent of $T_{th}$. From Eq.~(\ref{defTFD}), one has:
\be \label{Tth-TFD}
\frac{T_{th}}{T_{FD}} = 1 - \frac{\langle x \omega \rangle \, \langle g(x) \rangle }{\eta \,\langle x^2 \rangle}
\ee
where we have used the relations $T_{th}=\overline{\ve}/\eta$ and $\overline{\ve}=\langle g(x) \rangle$. The correlation $\langle x \omega \rangle$ can be written in a more explicit form as
\be \label{x-omega}
\langle x \omega \rangle = \left< \frac{x}{g'(x)} + (\eta-1) \, \frac{x^2}{g(x)} \right>
\ee
From Eqs.~(\ref{Tth-TFD}) and (\ref{x-omega}), it appears that the ratio
$T_{th}/T_{FD}$ generally depends on $T_{th}$, since the average
$\langle \dots \rangle$ is done with the one-site distribution which
is a function of temperature.

Now in the particular case where $g(x)$ is a power law, namely $g(x)=x^p/p$
(with $p$ an even integer), Eq.~(\ref{Tth-TFD}) actually simplifies to 
\be \label{TpropT}
T_{FD} = [2+p(\eta-1)] \, T_{th}
\ee
Note that for $2+p(\eta-1) \le 0$, the above equation would lead
to a negative $T_{FD}$, i.e.~a negative response $\chi(t)$ to the
perturbation $h$, which is rather counterintuitive.
Actually, $\chi(t)$ does not become negative in this case but diverges and
Eq.~(\ref{TpropT}) is no longer valid,
indicating the breakdown of linear response --the response is then non
linear with $h$ even for $h \to 0$.
This may be seen from the correlation $\langle x \omega \rangle$, which
can be written:
\be
\langle x \omega \rangle = A \int_0^{\infty} dx \, x^{1+p(\eta-1)}
e^{-x^p/pT_{th}}
\ee
where the constant $A$ depends on $p$ and $\eta$.
If $1+p(\eta-1) \le -1$ (i.e.~the same condition as above), the integral
diverges at its lower bound, and $\chi(t)$ becomes infinite.
To keep the susceptibility finite, one needs to consider values of $\eta$
such that $\eta > 1-2/p$.
It is interesting to note that as soon as $p>2$, the equilibrium
value $\eta = 1/p$ does not satisfy the above inequality, so that the
equilibrium response is non linear in this case. This is somehow
reminiscent of the Landau theory for phase transitions, in which the
magnetization $\langle m \rangle$ becomes non linear with the magnetic
field at the critical point, where the term in $m^2$ in the expansion of
the free-energy vanishes.

Finally, considering the specific case $g(x) = \frac{1}{2} x^2$ as
in \cite{court}, the above restriction disappears since $1-2/p=0$ for $p=2$.
The temperature $T_{FD}$ is then defined for all $\eta>0$
\footnote{Actually, only the behavior of $g(x)$ in the vicinity of $x=0$ is
responsible for the divergence of the susceptibility $\chi(t)$. For an even
regular function $g(x)$ such that $g''(0) \ne 0$, one has $g(x) \sim x^2$
for $x \to 0$, and the response remains linear in $h$ for all positive value
of $\eta$.}.
Using Eqs.~(\ref{eq-state}) and (\ref{TpropT}), one can write $T_{FD}$
in a very simple form which does not depend on $\eta$:
\be
T_{FD} = 2\overline{\ve}
\ee
where $\overline{\ve}$ is the energy density $\overline{\ve} = \frac{1}{2} \langle x^2
\rangle$.

To sum up, several different cases have to be distinguished.
For general regular functions $g(x)$ with
$g(x) \sim x^p$ for $x \to 0$, where $p>2$ an even integer, the response
is non linear with the field $h$ if $\eta \le 1-2/p$.
Otherwise, the response is linear and the susceptibility $\chi(t)$ can
be defined.
In this case, the FDR (or equivalently, the
fluctuation-dissipation plot) is generically non linear.
Now, several additional assumptions on $g(x)$ can be made:
if $g(x)$ is even, the FDR
is linear, leading to the definition of $T_{FD}$ as the inverse slope of
the FDR; yet, $T_{FD}$ is a priori not proportional to $T_{th}$.
Besides, if $g(x)$ is a power law (and if the response is
linear), then $T_{FD}$ becomes proportional to $T_{th}$.
The equality $T_{FD}=T_{th}$ is recovered only for $p=2$ and
$\eta=\frac{1}{2}$, i.e.~when linear response and microreversibility hold.

\subsection{Physical relevance of the different temperatures}
\label{sect-relevance}

In the preceding sections, two different temperatures have been
introduced:  a first one ($T_{th}$) from statistical considerations, and
a second one ($T_{FD}$) from a FDR.  These two temperatures do not only
have different definitions, but they also take different values, as seen
from Eq.~(\ref{Tth-TFD}).
In this section, we wish to compare the physical relevance of these two
definitions, and see whether or not both of them satisfy the basic
properties expected for a temperature.

\subsubsection{Inhomogeneous version of the model}

Considering a homogeneous system as we have done up to now, it is clear
that if $T_{th}$ takes the same value in two subsystems, so does $T_{FD}$
since the two temperatures are related through Eq.~(\ref{Tth-TFD}).
Indeed, if $g(x)=x^p/p$ these two temperatures are proportional according
to Eq.~(\ref{TpropT}), so that they may be considered to be identical
up to a redefinition of the temperature scale.
As a result, it seems not to be possible to discriminate between these
two definitions within the present model.

Actually, this apparent equivalence of both temperatures comes from the
fact that the parameter $\eta$ is the same throughout the system.  So
one could try to propose a generalization of the model in which $\eta$
would not be constant, still keeping the model tractable.  This can be
realized in the following way.  Introducing on each site $i$ a
parameter $\eta_i >0$, we define on each link $(j,k)$ a distribution
$\psi_{jk}(q)$ through
\be \label{psi-jk}
\psi_{jk}(q) =
\frac{\Gamma(\eta_j+\eta_k)}{\Gamma(\eta_j) \, \Gamma(\eta_k)} \,
q^{\eta_j-1} (1-q)^{\eta_k-1}
\ee
The redistribution rules are assumed
to keep the same form as in Eq.~(\ref{def-model}).  Yet, links $(j,k)$
now need to be oriented since $\psi_{jk}(q)$ is no longer symmetric, so
that the fraction $q$ is attributed to site $j$, whereas $1-q$ is
attributed to site $k$, precisely as in Eq.~(\ref{def-model}).

Note however that even though the redistribution process is locally
biased if $\eta_j \ne \eta_k$, there is no global energy flux in the
system since the form (\ref{psi-jk}) has been chosen to preserve the
detailed balance relation.  As a result, the steady-state distribution
can be computed exactly for any set of variables $\{\eta_i\}$.
To simplify the discussion, we restrict the results presented here
to the simple case $g(x)=\frac{1}{2} x^2$, but generalization to other
functions $g(x)$ are rather straightforward.
In this case, the `microcanonical' distribution $P(\{x_i\}|E)$ takes
essentially the same form as previously:
\be
P(\{x_i\}|E) = \frac{1}{\tilde{Z}_N(E)} \, \prod_{i=1}^N |x_i|^{2\eta_i-1} \delta \left( \frac{1}{2} \sum_{i=1}^N x_i^2 - E \right)
\ee
Following the same reasoning as above, one can define both $T_{th}$ and
$T_{FD}$ in this generalized model.  In particular, the temperature
$T_{th}^{\ell}$ is defined from the conditional probability $P(E_1|E)$
as in Eq.~(\ref{def-Tth-ell}).  Considering again a partition of a
large isolated system into two subsystems $\mathcal{S}_1$ and
$\mathcal{S}_2$, one finds for the subsystem $\mathcal{S}_{\ell}$
\be \label{TFDij}
T_{th}^{\ell} = \frac{\overline{\ve}_{\ell}}{\langle \eta \rangle_{\ell}} \qquad
T_{FD}^{\ell} = 2 \overline{\ve}_{\ell}
\ee
where $\langle \eta \rangle_{\ell}$ is the average of $\eta_i$ over the subsystem $\mathcal{S}_{\ell}$:
\be
\langle \eta \rangle_{\ell} \equiv \frac{1}{N_{\ell}} \sum_{i \in \mathcal{S}_{\ell}} \eta_i
\ee
If one chooses the set of variables $\{\eta_i\}$ such that $\langle
\eta \rangle_1 \ne \langle \eta \rangle_2$, the equality $T_{th}^1 =
T_{th}^2$, which is true from the very definition of $T_{th}^{\ell}$
--see Eq.~(\ref{eqTmodel})-- implies $\overline{\ve}_1 \ne \overline{\ve}_2$.  Consequently,
equipartition of energy breaks down, and from Eq.~(\ref{TFDij}) one has
$T_{FD}^1 \ne T_{FD}^2$: the fluctuation-dissipation temperature does
not take equal values in two subsystems \footnote{The same conclusions
hold for more general functions $g(x)$, but the results then take a less
concise form.}.

This last point is indeed reminiscent of recent numerical results
reported in the context of binary granular gases \cite{Barrat03}, where
the temperature associated to each species of grains from a FDR does
not equilibrate.  These results indicate that for {\it non glassy}
systems, the temperature defined from FDR does not fulfill the basic
properties required for a temperature, as the equality of the
temperatures of subsystems when a steady state has been reached.  On
the contrary, the temperature $T_{th}$ defined from statistical
considerations satisfies this property, and may thus be given a more
fundamental status.

Finally, it should be noticed that the relation $T_{th}=\overline{\ve}/\eta$
indicates that the temperature $T_{th}$ is not simply a measure of
the average energy, but also takes into account the fluctuations
of energy. Indeed, a large value of $\eta$ corresponds on the one hand
to a low value of the temperature, and on the other hand to a sharp
distribution $\psi(q)$, which in turn leads to small energy fluctuations
in the system, as can be seen for instance from the canonical distribution
given in Eq.~(\ref{eq-Pcan}).

\subsubsection{How to define a thermometer?}

Once a temperature has been formally defined in a system, a very
important issue is to be able to measure it, at least within a
conceptual experiment.  This question is in general highly non trivial
for out-of-equilibrium systems.  In the context of glassy systems for
instance, it has been proposed to use a simple harmonic oscillator
connected to the system as a thermometer \cite{CuKuPe}.  Still, in
order to measure a temperature associated to a given time scale $\tau$
(assumed to be large with respect to the microscopic time scale
$\tau_0$), one must use an harmonic oscillator with a characteristic
time scale of the order of $\tau$.  In this case, the temperature is
obtained through the usual relation $\overline{\ve}_{osc} = \frac{1}{2} T$, where
$\overline{\ve}_{osc}$ is the average kinetic energy of the oscillator.  For
glassy systems, this temperature has also been shown to identify with
the temperature defined from FDR \cite{CuKuPe}.  Besides, a numerical
realization of such a thermometer has been proposed by using a brownian
particle with a mass much larger than the other particles, in a glassy
Lennard-Jones mixture under shear \cite{Berthier-thermo}.  Such a
definition of temperature is also consistent with the so-called
`granular temperature', defined as $2/d$ times the average kinetic
energy of the grains \cite{Haff} ($d$ is the space dimension).

Interestingly, in the present model which is not glassy, a somewhat analogous
procedure would be to connect a new site to the system, and make it
interact with the other sites using the current kinetic rules of the
model; this new site would play the role of a thermometer.  Assuming
again $g(x)=\frac{1}{2} x^2$, the temperature read off from the
average energy of the thermometer is precisely $T_{FD}$. At first
sight, this seems to be in contradiction with the above discussion
in which we argued that $T_{th}$ was the physically relevant
temperature. The paradox comes from the fact that we used without
justifying it the relation $\overline{\ve}_{osc} = \frac{1}{2} T$
to define the
temperature $T$ of the thermometer as a function of the measurable quantity
$\overline{\ve}_{osc}$.
Accordingly, such a definition does not ensure that $T$
is the temperature of the system.

One of the most important properties of $T_{th}$ is precisely that it
takes equal values within subsystems in contact. Actually, to obtain
$T_{th}$, one needs to know the equation of state of the thermometer,
that relates measurable quantities like the average energy $\overline{\ve}_{osc}$
to the temperature $T_{th}$.  Indeed, the fact that it is necessary to
know the equation of state of the thermometer in order to measure the
temperature is not a specificity of non-equilibrium states, but is
also true in equilibrium situations, in which one must know for
instance the relation between the height of a liquid in a vertical
pipe and the temperature of this liquid.  In the same way, the
relation $\overline{\ve}_{osc} = \frac{1}{2} T$ invoked
above is not obvious in
itself, but results from equilibrium statistical mechanics.  As a
result, there is no clear reason why this last relation should hold
for generic non-equilibrium situations.

Yet, an important point must be mentioned at this stage.  One of the
specificity of non-equilibrium states is that there is not a unique way
to define a thermal contact between two systems.  In equilibrium, it is
usually enough to consider the weak interaction limit in which the energy
associated with the interaction process is very small compared to the
other energies involved.  On the contrary, for non-equilibrium systems,
the conservation of energy is not sufficient, since the dynamics
can be much richer, as illustrated by the presence of the parameter
$\eta$ in the present model. Hereabove, we assumed that the new site
used as a thermometer was driven by the same dynamical rules as the system
it is in contact with. Yet, in practical situation, one would rather use
a thermometer with a known equation of state to measure the temperature
of another system for which the equation of state is {\it unknown}.
As a consequence, the dynamics of the thermometer is expected in general to be
different from that of the system. Determining the properties
that a thermometer has to satisfy in order to measure correctly the
temperature thus remains an open question.

\section{Renormalization approach}

\subsection{Breaking of detailed balance}

If $\psi(q)$ is different from a beta law, no simple detailed balance
relation has been found in this model.  In the absence of such a
relation, it is rather hopeless to find the stationary distribution
$P_{st}(\{x_i\}|E)$, even though some sophisticated algebraic methods have
proven to be efficient in some cases \cite{Derrida,Sandow}.  Yet, the
fact that we were not able to find a detailed balance relation in the
model is not a proof that the relation does not exist.  As a result, it
appears  useful to test numerically the existence of non zero
probability fluxes even in steady state, which would clearly
demonstrate the absence of detailed balance.

\begin{figure}[t]
\centering\includegraphics[width=7.5cm,clip]{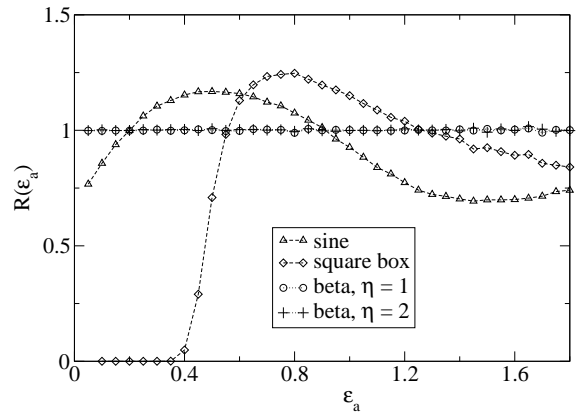}
\caption{\sl Ratio $R(\ve_a)$ of forward and backward probabilities of a
transition path; $R(\ve_a) \ne 1$ indicates a breaking of detailed balance.
Distributions $\psi(q)$ used are the sine-like distribution ($\vartriangle$)
and the `square box' one ($\diamond$) --see text for details.
Results for beta $\psi(q)$ with $\eta=1$ ($\circ$) and $\eta=2$ ($+$) are
also presented for comparison, showing as expected that detailed balance is
satisfied for these distributions.}
\label{fig-DB}
\end{figure}

As discussed in Sect.~\ref{sect-steady}, the steady-state distribution
can be fully determined in terms of the dynamics of the local energy
$\ve_i = g(x_i)$. In the following, we thus use these variables
$\ve_i$ as the dynamical variables. The dynamics of $\ve_i$ is the
same as that of the variables $x_i$ if one considers the case $g(x_i)=x_i$,
restricting $x_i$ to be positive. The detailed balance property is
checked by measuring with numerical simulations the probability
$p_{ab}(\ve_a,\ve_b,\delta\ve)$ to observe on a given site $i$ a direct
transition from a value $\ve_i \in [\ve_a, \ve_a + \delta\ve]$ to a
new value $\ve_i' \in [\ve_b, \ve_b + \delta\ve]$, as well as the
reverse probability $p_{ba}(\ve_b,\ve_a,\delta\ve)$ to go from the
interval $[\ve_b, \ve_b + \delta\ve]$ to the interval $[\ve_a, \ve_a +
\delta\ve]$.  These probabilities are actually obtained by averaging
over all sites $i$.  One then computes the ratio
\be
R = \frac{p_{ab}(\ve_a,\ve_b,\delta\ve)}{p_{ba}(\ve_b,\ve_a,\delta\ve)}
\ee
which becomes independent of $\delta\ve$ in the limit of small
$\delta\ve$.  Besides, a simple parametrization is to set $\ve_b =
\ve_a+\Delta$, and to compute $R$ as a function of $\ve_a$ for a fixed
value of $\Delta$.  Fig.~\ref{fig-DB} presents the numerical results
obtained for $R(\ve_a)$ with $\Delta=E/N$, using distributions
$\psi(q)$ which differ significantly from beta laws as the
sine-like distribution $\psi(q)=\frac{\pi}{2} |\sin(2\pi q)|$, and the
`square box' one, $\psi(q)=2$ for $\frac{1}{4} < q < \frac{3}{4}$, and
$\psi(q)=0$ otherwise.  Beta laws are also shown for comparison.  As
expected, $R(\ve_a)=1$ for beta laws, whereas $R(\ve_a) \ne 1$ for
other distributions, showing that detailed balance is broken in this
case.

\subsection{Numerical renormalization procedure}

Even though detailed balance is broken microscopically when $\psi(q)$
is different from a beta law, one can wonder whether the macroscopic
properties of the model differ significantly or not from that in the
presence of detailed balance.  Indeed, some studies
\cite{Grinstein,Tauber} have shown that a weak breaking of detailed
balance does not influence the critical properties of particular
classes of spin models.  In the present model, numerical simulations
suggest that even for distributions $\psi(q)$ with a behavior far from
beta laws, no spatial correlations appear within two-points functions.
Note that this result is also consistent with the vanishing of
two-point correlations in the `q-model' for granular matter
\cite{Snoeijer}, which presents some formal similarities (although in a
different spirit), but also important differences, with the present
model. In particular, the q-model is static, and the role played by time
here corresponds to the vertical space direction.  In addition, the
dynamics of the q-model is equivalent to a synchronous dynamics, and
the conserved quantity is linear since it represents the vertical
component of forces between grains.

In order to test whether macroscopic properties are influenced or not
by the breaking of detailed balance at the microscopic level, one can
try to use a renormalization group approach.  Even though such an approach
might not seem natural in a context where no diverging length scale
appears, this is actually a standard way to compute the effective dynamics at
a coarse-grained level.  Since no analytical solution is available for
$\psi(q)$ different from a beta law, one has to resort to numerical
simulations.

To this aim, the following renormalization procedure is introduced.
The $d$-dimensional lattice is divided into cells (or blocks) of linear
size $L$, and the effective dynamics between cells is measured from
numerical simulations of the microscopic dynamics.  To be more
specific, when running the microscopic dynamics, one has to choose at
random a link of the lattice at each time step, and to redistribute the
energy over the link.  If both sites of this link belong to the same
block, then the redistribution is only an intra-block dynamics, and
corresponds precisely to the degrees of freedom that have to be
integrated out by the renormalization procedure.  As a result, nothing
is recorded during this particular process.

On the contrary, if the chosen link lies between two different cells,
then the process is considered as a redistribution between blocks, and
the effective fraction $q_R$ of energy redistributed is computed. 
Having chosen an
orientation of the lattice, one can label for instance by $1$ and $2$
the two blocks involved in the process. Clearly, the total energy of
these two blocks is conserved during this process.  One thus computes
the energy $E_b^1$ and $E_b^2$ of each block, and defines the effective
fraction $q_R$ as the ratio:
\be
q_R = \frac{E_b^1}{E_b^1+E_b^2}
\ee
To obtain the renormalized energy, one should actually divide $E_b$
by the size of the block (so that the energy density is conserved),
but this is not essential here since we consider only energy ratios.
The histogram of the values of $q_R$ obtained when
running the microscopic dynamics is recorded, which gives the renormalized
distribution $\psi_L(q)$.  One would like to test if for large values
of $L$, detailed balance is recovered, which would mean that the
distribution $\psi_L(q)$ converges (in some sense to be specified)
towards a beta law.  As usual with renormalization procedures, the
correct way to obtain large block sizes is not to consider large blocks
from the beginning, but instead to start from small blocks and to
iterate the procedure until the desired size is reached.

\begin{figure}[t]
\centering\includegraphics[width=7.5cm,clip]{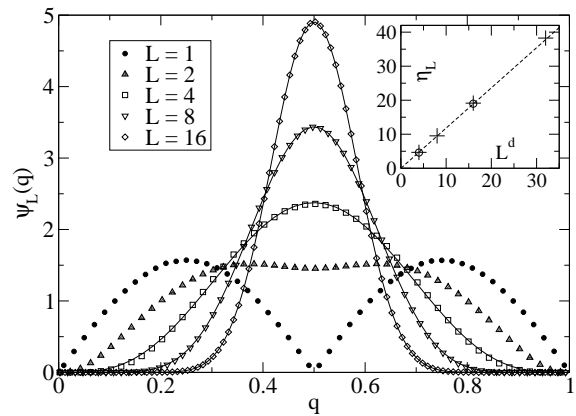}
\caption{\sl Renormalized distribution $\psi_L(q)$ for increasing sizes $L$,
in dimension $d=1$. Full lines correspond to one-parameter fits with beta
distributions. Inset: parameter $\eta_L$ from the fit plotted as a function
of $L^d$ for $d=1$ ($+$) and $d=2$ ($\circ$); dashed line is the mean-field
prediction given in Eq.~(\protect{\ref{eq-eta_e}}).
}
\label{fig-renor}
\end{figure}

As a result, we started from cells of size $L=2$ and computed
successively $\psi_2(q)$, $\psi_4(q)$, $\psi_8(q)$, etc., by applying
recursively the same procedure with a microscopic dynamics defined by
the renormalized $\psi_L(q)$ obtained at the step before.  Numerical
results obtained starting from an initial distribution $\psi(q) =
\frac{\pi}{2} |\sin(2\pi q)|$ are shown on Fig.~\ref{fig-renor}, for
space dimensions $d=1$ and $d=2$. For $L \ge 4$, the resulting
distributions $\psi_L(q)$ can be very well fitted by beta laws, i.e.~by a
test distribution $\psi_{\rm test}(q)$ of the form
\be
\psi_{\rm test}(q) = \frac{\Gamma(2\eta_L)}{\Gamma(\eta_L)^2} \, [q (1-q)]^{\eta_L-1}
\ee
with only one free parameter $\eta_L$.  This parameter $\eta_L$ is an
increasing function of $L$, which can be easily understood from the
fact that increasing the size of the blocks reduces the fluctuations of
the energy from one block to another.  So if one lets the size $L$ go
to infinity, the distribution $\psi_L(q)$ eventually converges to a
Dirac delta function centered on $q=\frac{1}{2}$.  This means that the
beta laws found from fitting the data are to be understood as
pre-asymptotic distributions rather than as true limit distributions.

Very interestingly, the fitting parameter $\eta_L$ is found to be
linear with $L^d$, as seen in the inset of Fig.~\ref{fig-renor}.  This
behavior can be interpreted in the following way, assuming that the
initial distribution $\psi(q)$ is a beta law with parameter $\eta$.  As
seen from the calculations done in Sect.~\ref{sect-steady}, the local
distribution of the energy $\varepsilon_i=g(x_i)$ is given by a gamma
law of exponent $\eta$ and scale parameter $\beta=1/T_{th}$:
\be \label{p-gamma}
p(\varepsilon_i) = \frac{\beta^\eta}{\Gamma(\eta)} \, \varepsilon_i^{\eta-1} e^{-\beta \varepsilon_i}
\ee
The $\varepsilon_i$'s are independent random variables, so that the block energies $E_b$, defined as:
\be \label{eq-Eblock}
E_b = \sum_{i \in {\rm block}} \varepsilon_i
\ee
are distributed according to beta laws with exponent $\eta L^d$, where
$L^d$ is the number of sites within a block.  Then taking the ratio
$q_R=E_b^1/(E_b^1+E_b^2)$, one obtains for $q_R$ a beta distribution of
parameter $\eta L^d$, as is well-known from the properties of gamma
laws.

So starting from a beta law for $\psi(q)$, the above analytical
argument shows that beta laws are again obtained from the
renormalization procedure, with a parameter $\eta_L$ linear in $L^d$.
Interestingly, the coefficient of proportionality is precisely the
parameter $\eta$ of the microscopic law $\psi(q)$.  So when starting
from an arbitrary distribution $\psi(q)$, it is natural to define an
effective parameter $\eta_e$ from the fitting parameter $\eta_L$ as:
\be
\eta_e = \frac{\eta_L}{L^d}
\ee
One can then interpret $\eta_e$ as the parameter of the microscopic
beta law which would give the same macroscopic behavior of the system
as the initial distribution $\psi(q)$.

\subsection{Mean-field predictions}
\label{sect-MF}

In this section, we aim to predict within a mean-field framework the
effective exponent $\eta_e$ introduced above, for an arbitrary
distribution $\psi(q)$.  In a mean-field description, one assumes that
the two-site steady-state distribution $P_2(x_1,x_2)$ can be factorized
as a product of one-site distributions:
\be
P_2(x_1,x_2) = P_1(x_1) \, P_1(x_2)
\ee
This assumption is valid if $\psi(q)$ is a beta law, as can be seen
from Eq.~(\ref{eq-Pcan}). For more general $\psi(q)$, it remains a
priori only an approximation.  In order to deal with the
renormalization procedure, it is more convenient to work with the
distribution $p(\varepsilon_i)$ of the local energy $\varepsilon_i
\equiv g(x_i)$, rather than with $P_1(x_i)$.
In terms of the variables $\{\ve_i\}$, the redistribution rules read:
\be
\varepsilon_j' = q\, (\varepsilon_j + \varepsilon_k), \qquad
\varepsilon_k' = (1-q)\, (\varepsilon_j + \varepsilon_k)
\ee
Numerical simulations show that after a sufficient coarse-graining by
the renormalization procedure, the renormalized distribution
$\psi_L(q)$ becomes a beta law with parameter $\eta_L = \eta_e \,
L^d$.  The associated renormalized distribution of the block energies
$E_b$ is then a gamma law with exponent $\eta_L$ and scale parameter
$\beta_L$:
\be
p_L(E_b) = \frac{\beta_L^{\eta_L}}{\Gamma(\eta_L)} \, E_b^{\eta_L-1} e^{-\beta_L E_b}
\ee
The exponent $\eta_L$ can be determined from the first and second
moments of the distribution $p_L(E_b)$. Indeed, one finds an average
value $\langle E_b \rangle = \eta_L/\beta_L$, and a variance ${\rm
Var}(E_b) = \eta_L/\beta_L^2$, with ${\rm Var}(E_b) = \langle E_b^2
\rangle - \langle E_b \rangle^2$. As a result, $\eta_L$ is given by:
\be \label{eq-etaL}
\eta_L = \frac{\langle E_b \rangle^2}{\langle E_b^2
\rangle - \langle E_b \rangle^2}
\ee
If the initial distribution $p(\varepsilon_i)$ is factorized, the
block energies $E_b$ are sums of independent random variables --see
Eq.~(\ref{eq-Eblock}).  So the average value and the variance of $E_b$
are simply the sums of the average and variance of the variables
$\varepsilon_i$:
\be
\langle E_b \rangle = \langle \varepsilon \rangle \, L^d, \qquad {\rm Var}(E_b) = {\rm Var}(\varepsilon) \, L^d
\ee
From Eq.~(\ref{eq-etaL}), the effective exponent $\eta_e = \eta_L/L^d$ is
thus found to be:
\be
\eta_e = \frac{\langle \varepsilon \rangle^2}{{\rm Var}(\varepsilon)}
\ee
So if we know the two first moments of the distribution
$p(\varepsilon)$, we are able to compute $\eta_e$.

To obtain these moments for an arbitrary $\psi(q)$, we use the
following steady-state master equation for the distribution
$p(\varepsilon)$:
\bea \nonumber
p(\varepsilon) &=& \int_0^{\infty} d\varepsilon_1 \, p(\varepsilon_1) \int_0^{\infty} d\varepsilon_2 \, p(\varepsilon_2) \times\\
&& \quad \int_0^1 dq \, \psi(q) \, \delta \left( \varepsilon - q(\varepsilon_1 + \varepsilon_2) \right)
\label{EM-MF}
\eea
This equation can be considered as describing the redistribution
process over an isolated single link.  Yet, it can also be derived from
a mean-field version of the model, in which redistributions can occur
over any pair of sites of the system --see Appendix~\ref{apMF}.

Introducing the Laplace transform $\hat{p}(s)$ defined as
\be
\hat{p}(s) \equiv \int_0^{\infty} d\varepsilon \, e^{-s\varepsilon} \, p(\varepsilon)
\ee
one can rewrite Eq.~(\ref{EM-MF}) as
\be
\hat{p}(s) = \int_0^1 dq \, \psi(q) \int_0^{\infty} d\varepsilon_1 \, p(\varepsilon_1) \int_0^{\infty} d\varepsilon_2 \, p(\varepsilon_2) \, e^{-sq(\varepsilon_1+\varepsilon_2)}
\ee
The integrals over $\varepsilon_1$ and $\varepsilon_2$ can be
factorized into a product of Laplace transforms:
\be \label{EM-MF-TL}
\hat{p}(s) = \int_0^1 dq \, \psi(q) \, \hat{p}(qs)^2
\ee
From the last equation, the successive moments of $p(\varepsilon)$ can be obtained, since they are given by the derivatives of $\hat{p}(s)$ in $s=0$:
\be
\langle \varepsilon \rangle = - \frac{d\hat{p}}{ds} \Big\vert_{s=0}, \qquad
\langle \varepsilon^2 \rangle = \frac{d^2\hat{p}}{ds^2} \Big\vert_{s=0}
\ee
Note also that by definition, $\hat{p}(0)=1$.
Taking the first derivative of Eq.~(\ref{EM-MF-TL}) in $s=0$, one recovers that $\langle q \rangle = \frac{1}{2}$.
More interestingly, the second derivative of Eq.~(\ref{EM-MF-TL}) yields:
\be
\frac{d^2\hat{p}}{ds^2} \Big\vert_{s=0} = 2 \int_0^1 dq \, q^2 \, \psi(q) \left[ \left( \frac{d\hat{p}}{ds} \Big\vert_{s=0} \right)^2 + \frac{d^2\hat{p}}{ds^2} \Big\vert_{s=0} \right]
\ee
In terms of moments, the last equation reads:
\be
\langle \varepsilon^2 \rangle = 2 \langle q^2 \rangle \left[ \langle \varepsilon \rangle^2 + \langle \varepsilon^2 \rangle \right]
\ee
To compute $\eta_e$, we only need the ratio $\langle \varepsilon \rangle^2/\langle \varepsilon^2 \rangle$, which is easily found from the preceding equation:
\be
\frac{\langle \varepsilon^2 \rangle}{\langle \varepsilon \rangle^2} = \frac{2 \langle q^2 \rangle}{1-2\langle q^2 \rangle}
\ee
Taking into account that $\langle q \rangle = \frac{1}{2}$, $\eta_e$ is found to be:
\be \label{eq-eta_e}
\eta_e = \frac{1}{8 {\rm Var}(q)} - \frac{1}{2}
\ee

\subsection{Analytical arguments}

To conclude this section dedicated to renormalization group approaches,
we wish to give a heuristic analytical argument that may help to 
understand the numerical results presented on Fig.~\ref{fig-renor}.
As explained above, the renormalization can be worked out
exactly in the case where $\psi(q)$ is a beta law. The numerical
procedure shows that other distributions $\psi(q)$
converge to beta laws under renormalization.
From an analytical point of view, it is more convenient to work with
the distribution $p(\ve)$ of the local energy, rather than with $\psi(q)$.
A beta $\psi(q)$ is associated to a gamma law for $p(\ve)$ so that it
would be interesting to check analytically whether an arbitrary $p(\ve)$
converges to a gamma law under renormalization.
Note that an implicit assumption here
is that the $N$-site energy distribution is factorized, in a mean-field
spirit. 

A general calculation for an arbitrary initial distribution $p(\ve)$ is
in fact highly non trivial. We thus restrict the following calculations
to an initial $p(\ve)$ which differs only slightly from a gamma law:
\be \label{pve-perturb}
p(\ve) = p_{\gamma}(\ve) + \lambda \delta p(\ve)
\ee
where $\lambda \ll 1$ is an arbitrarily small parameter, and
$p_{\gamma}(\ve)$ is a gamma distribution similar to that used in
Eq.~(\ref{p-gamma}). Since the renormalization conserves the average
energy, $p(\ve)$ and $p_{\gamma}(\ve)$ must have the same average value
$\overline{\ve}$ so as to become equivalent after renormalization.
Taking also into account the normalization condition, $\delta p(\ve)$
has to satisfy
\be \label{moment01}
\int_0^{\infty} d\ve \, \delta p(\ve) = 0, \qquad
\int_0^{\infty} d\ve \, \ve \,\delta p(\ve) = 0.
\ee
Let $M \equiv L^d$ be the number of sites in a block. The renormalized energy
$\ve_R$ is given by
\be
\ve_R = \frac{1}{M} \sum_{i \in {\rm block}} \ve_i
\ee
The distribution of $p_1(\ve_R)$ is more easily obtained using a Laplace
transform:
\be \label{eq-p1p-renor}
\hat{p_1}(s) = \hat{p}(s/M)^M
\ee
Obviously, a fixed point for this equation is $p(s)=e^{-s \overline{\ve}}$,
which leads to $p(\ve) = \delta(\ve-\overline{\ve})$. The aim of the present
calculation is to see whether $p(\ve)$ and $p_{\gamma}(\ve)$ converge
`in the same way' or not toward the delta distribution.

Replacing Eq.~(\ref{pve-perturb}) into Eq.~(\ref{eq-p1p-renor}) and
expanding up to first order in $\lambda$, one has:
\be
\delta \hat{p}_1(s) = M \, \hat{p}_{\gamma}\left(\frac{s}{M}\right)^{M-1}
\delta \hat{p}\left(\frac{s}{M}\right)
\ee
Iterating $K$ times the renormalization procedure, one gets:
\be \label{delta-ps}
\delta \hat{p}_K(s) = M^K  \, \delta \hat{p}\left(\frac{s}{M^K}\right)
\prod_{n=0}^{K-1} \hat{p}_{\gamma,n}\left(\frac{s}{M^{K-n}}\right)^{M-1}
\ee
The renormalized gamma distribution $\hat{p}_{\gamma,n}(s)$ 
obtained after $n$ iterations is given by
\be
\hat{p}_{\gamma,n}(s) =
\left( 1+\frac{s\overline{\ve}}{\eta \, M^n} \right)^{-\eta M^n}
\ee
Then Eq.~(\ref{delta-ps}) can then be rewritten:
\be
\delta \hat{p}_K(s) = M^K  \, \left( 1+\frac{s \overline{\ve}}{\eta \, M^K}
\right)^{-\eta(M^K-1)} \, \delta \hat{p} \left(\frac{s}{M^K}\right)
\ee
Using the relation
\be
\lim_{K \to \infty} \left( 1+\frac{s \overline{\ve}}{\eta \, M^K}
\right)^{-\eta(M^K-1)}
= e^{-s \overline{\ve}}
\ee
one ends up with
\be
\delta \hat{p}_K(s) \approx M^K \, e^{-s \overline{\ve}}
\, \delta \hat{p} \left(\frac{s}{M^K}\right)
\ee
Expanding $\delta \hat{p}(s)$ in power of $s$ for $s \to 0$, one has
\be
\delta \hat{p}(s) = g_2 s^2 + {\cal O}(s^3)
\ee
since the terms of order $0$ and $1$ vanish due to Eq.~(\ref{moment01}).
This yields:
\be
\delta \hat{p}_K(s) \approx e^{-s \overline{\ve}} \, \frac{g_2 s^2}{M^K}
\ee
which goes to $0$ when $K \to \infty$ as expected. Yet, this is not enough to
show that $p(\ve)$ and $p_{\gamma}(\ve)$ converge `in the same way' toward
the distribution $\delta(\ve-\overline{\ve})$. To do so, one has to show
that $\delta \hat{p}_K(s)$ goes to $0$ more rapidly than the `distance'
between $\hat{p}_{\gamma,K}(s)$ and the infinite $K$ limit
\be
\hat{p}_{\gamma,\infty}(s) = e^{-s\overline{\ve}}
\ee
A way to quantify this `distance' is to introduce the quantity:
\be
D_K = \int_0^{\infty} ds \, \big\vert \hat{p}_{\gamma,K}(s) - \hat{p}_{\gamma,\infty}(s) \big\vert
\ee
which can be shown easily to take the asymptotic form:
\be
D_K \approx \frac{1}{\eta \, \overline{\ve} M^K}
\ee
The convergence criterion can be written as
\be
\lim_{K \to \infty} \frac{\delta \hat{p}_K(s)}{D_K} = 0
\ee
This requires that $g_2=0$ in the expansion of $\delta \hat{p}_0(s)$,
which implies that the distributions $p(\ve)$ and $p_{\gamma}(\ve)$ have
the same variance $\sigma^2=\sigma_{\gamma}^2$. Such a condition is
actually natural, as the variance becomes $\sigma_K^2=\sigma^2/M^K$ under
renormalization. If the two distributions take the same form after
renormalization, they should have in particular the same variance
$\sigma_K^2 = \sigma_{\gamma,K}^2$, and one recovers
$\sigma^2=\sigma_{\gamma}^2$.

Obviously, the above arguments are not fully rigorous, and remain somehow
at a heuristic level, but they already give some insights on the mechanisms
leading to the convergence process observed numerically.

\section{Conclusion}

The class of models studied in the present paper is a very
interesting example in which one can define a meaningful
temperature $T_{th}$ from the conditional energy distribution of two
subsystems, a procedure similar to the one used in the equilibrium
microcanonical ensemble.
These models exhibit a rich behavior which includes linear as well as
non linear response to a perturbation, and linear or non linear
fluctuation-dissipation relations when the response is linear.
Our major result is that the temperature
$T_{FD}$ deduced from the (linear) FDR does not coincide with the statistical
temperature $T_{th}$, and that $T_{FD}$ does not take equal values in
two subsystems when one considers an inhomogenous version of the
model.  This suggests that FDR are not necessarily the relevant way to
define a temperature in the context of non glassy out-of-equilibrium
steady-state systems.

In addition, a numerical renormalization procedure suggests that
detailed balance is generically restored on a coarse-grained level when
it is not satisfied by the microscopic dynamics.  This
renormalization procedure yields a new parameter $\eta_e$ describing
the deviation from equilibrium, which can be analytically computed
within a mean-field approximation.  This leads to a macroscopic
description of the system with two parameters, namely $T_{th}$ and
$\eta_e$.

Finally, from a more general point of view, the present work raises
important questions concerning the way to extend the concepts of
statistical mechanics and thermodynamics to out-of-equilibrium systems.
On the one hand, the very definition of thermometers in non-equilibrium
systems appears to be a highly non trivial issue, as
the way to couple the thermometer to the system is not
unique. Thus one may need to impose some --still unknown--
prescriptions on the coupling to get a well-defined measurement. On
the other hand, the present work may be of some relevance for the
description of non-equilibrium systems in which a global quantity is
conserved. For instance, one may think of the two-dimensional
turbulence where the vorticity is globally conserved
\cite{Kraichnan,Lesieur,Frisch},
or of dense granular matter in a container with fixed volume, in which
the sum of the local free volumes would also be conserved. Indeed, the
present model, for which the probability distribution is generically
non uniform over the mutually accessible states (i.e.~states with the
same value of the energy, or volume, etc.) may allow in particular to
go beyond the so-called Edwards' hypotheses \cite{Edwards,Mehta,Barrat00},
according to which all
accessible blocked states have the same probability to be occupied.

\subsection*{Acknowledgements}

E.B.~is particularly grateful to J.-P.~Bouchaud and F.~Lequeux for important
contributions which were at the root of the present work.
This work has been partially supported by the Swiss National Science
Foundation.

\appendix

\section{Time-dependent entropy}
\label{ap-entropy}

In this appendix, we show that the time-dependent entropy $S_E(t)$
defined in Eq.~(\ref{SEt-def}) is a non-decreasing function of time.
Let us first recall its definition:
\be
S_E(t) = - \int \prod_{i=1}^N dx_i \, P(\{x_i\},t) \,
\ln \frac{P_E(\{x_i\},t)}{f(\{x_i\})}
\ee
with $f(\{x_i\}) \equiv \prod_{i=1}^N |g'(x_i)| \, g(x_i)^{\eta-1}$,
and $P_E(\{x_i\},t)$ is the probability measure restricted to the
hypersurface of given energy $E$:
\be
P(\{x_i\},t) = P_E(\{x_i\},t) \; \delta \left( \sum_{i=1}^N g(x_i)-E \right)
\ee
Taking the derivative with respect to time, one finds:
\be
\frac{d S_E}{dt} = - \int \prod_{i=1}^N dx_i \, \frac{\partial P}{\partial t}(\{x_i\},t) \, \ln \frac{P_E(\{x_i\},t)}{f(\{x_i\})}
\ee
since the integral of the time derivative of the logarithm vanishes.
One can then use the master equation to express $\partial P/\partial t$
as a function of $P(\{x_i\},t)$ and of the transition rates. The
obtained expression can be symmetrized by permuting the integration
variables $x_i$ and $x_i'$ to get:
\bea \nonumber
\frac{dS}{dt} &=& \frac{1}{2} \int \prod_{i=1}^N dx_i \, dx_i' \, \left[ W(\{x_i'\}|\{x_i\}) \, P(\{x_i\},t) - \right. \\ \nonumber
&& \qquad \qquad \qquad \left. W(\{x_i\}|\{x_i'\}) \, P(\{x_i'\},t) \right] \, \times \\
&& \quad \qquad \left[ \ln \frac{P_E(\{x_i\},t)}{f(\{x_i\})} -  \ln \frac{P_E(\{x_i'\},t)}{f(\{x_i'\})} \right]
\eea
Now one can use the detailed balance relation Eq.~(\ref{eq-DB}):
\be
W(\{x_i'\}|\{x_i\}) \, f(\{x_i\}) = W(\{x_i\}|\{x_i'\}) \, f(\{x_i'\})
\ee
and write $dS/dt$ in the following way:
\bea \nonumber
\frac{dS}{dt} &=& \frac{1}{2} \int \prod_{i=1}^N dx_i \, dx_i' \, W(\{x_i'\}|\{x_i\}) \, f(\{x_i\}) \, \times \\ \nonumber
&& \quad \left[ \frac{P_E(\{x_i\},t)}{f(\{x_i\})} - \frac{P_E(\{x_i'\},t)}{f(\{x_i'\})} \right] \, \times \\
&& \quad \left[ \ln \frac{P_E(\{x_i\},t)}{f(\{x_i\})} -  \ln \frac{P_E(\{x_i'\},t)}{f(\{x_i'\})} \right]
\eea
In this form, it is clear that the time derivative of the entropy is always positive. It vanishes only for the steady state distribution:
\be
P_E(\{x_i\}) = \frac{1}{Z_N(E)} \, f(\{x_i\})
\ee
and the corresponding maximum value of the entropy is equal to:
\be
S(E) = \ln Z_N(E)
\ee

\section{Mean-field master equation}
\label{apMF}

In section \ref{sect-MF}, a simple steady-state master equation was
introduced to describe the one-site distribution $p(\ve_i)$ of the
energy $\ve_i \equiv g(x_i)$ --see Eq.~(\ref{EM-MF})-- in the case of
an arbitrary distribution $\psi(q)$.  We show here how this simple
equation can be derived from the master equation associated to a
$N$-site model with infinite range interactions.  Introducing such
long range interactions is a usual way to build a mean-field version of a
model.  To be more specific, we generalize the model introduced in
Eq.~(\ref{def-model}) in order to allow redistributions over any pair
of sites $(j,k)$, and not only on the links of the lattice.  As a
result, the lattice becomes useless in this version of the model.

The transition rates read:
\bea \label{rate-MF}
&& W(\{\ve_i'\}|\{\ve_i\}) = \frac{1}{N} \sum_{j<k} \left[ \prod_{i \ne j,k} \delta(\ve_i'-\ve_i) \right] \times \\ \nonumber
&& \quad  \delta(\ve_j'+\ve_k'-\ve_j-\ve_k) \int_0^1 dq \, \psi(q) \, \delta \left(\ve_j' - q(\ve_j+\ve_k) \right)
\eea
where the sum runs over all pairs $(j,k)$. The factor $1/N$ is
introduced so that each site keeps, in the thermodynamic limit $N \to
\infty$, a probability per unit time of the order of one to
be involved in a redistribution.

The stationary distribution $P_{MF}(\{\ve_i\})$ satisfies the following master equation:
\bea
&&P_{MF}(\{\ve_i\}) \int \prod_{i=1}^N d\ve_i' \, W(\{\ve_i'\}|\{\ve_i\}) = \\ \nonumber
&& \quad \int \prod_{i=1}^N d\ve_i' \, W(\{\ve_i\}|\{\ve_i'\}) \, P_{MF}(\{\ve_i'\})
\eea
The first integral is the total exit rate from configuration
$\{\ve_i\}$, and is equal to $(N-1)/2$ from Eq.~(\ref{rate-MF}).  So
the last equation can be rewritten in a more explicit form:
\bea \nonumber
&& P_{MF}(\{\ve_i\}) = \frac{2}{N(N-1)} \times \\
&& \quad \sum_{j<k} \int_0^{\infty} d\ve_j' \int_0^{\infty} d\ve_k' \, \delta(\ve_j'+\ve_k'-\ve_j-\ve_k) \times \\ \nonumber
&& \quad \int_0^1 dq \, \psi(q) \, \delta \left( \ve_j - q(\ve_j'+\ve_k') \right) P_{MF}(\ve_j',\ve_k',\{\ve_i\}_{i\ne j,k})
\eea
In order to go further, one has to assume that the distribution 
$P_{MF}(\{\ve_i\})$ factorizes:
\be
P_{MF}(\{\ve_i\}) = \prod_{i=1}^N p(\ve_i)
\ee
where $p(\ve)$ is the one-site distribution.
This assumption is justified in the limit of large $N$.
Integrating over all variables except $\ve_1$, one gets:
\bea
&& p(\ve_1) = \sum_{j=1}^{N-1} \sum_{k=j+1}^N \frac{2}{N(N-1)} \times \\ \nonumber
&& \int \prod_{i=2}^N d\ve_i \, \prod_{i \ne j,k} p(\ve_i) \int_0^{\infty} d\ve_j' \, p(\ve_j') \int_0^{\infty} d\ve_k' \, p(\ve_k') \times \\ \nonumber
&& \delta(\ve_j'+\ve_k'-\ve_j-\ve_k) \int_0^1 dq \, \psi(q) \, \delta \left(\ve_j - q(\ve_j'+\ve_k') \right)
\eea
The r.h.s.~can then be decomposed into two terms, one corresponding to
$i=1$ and the other one to $i>1$, which are called respectively $R_1$
and $R_2$ in the following:
\be \label{eq-R1R2}
p(\ve_1) = R_1 + R_2
\ee
The first term $R_1$ is associated with redistributions involving site
$j=1$ as well as another arbitrary site $k$.  It is actually
independent of $k$, so that $R_1$ is the sum of $(N-1)$ identical
terms.  Integrating over $\ve_k$ removes the delta distribution
$\delta(\ve_1'+\ve_k'-\ve_1-\ve_k)$, and one finds:
\bea
R_1 &=& \frac{2}{N} \int_0^{\infty} d\ve_1' \, p(\ve_1') \int_0^{\infty} d\ve_2' \, p(\ve_2') \times \\ \nonumber
&& \qquad \qquad \int_0^1 dq \, \psi(q) \, \delta \left(\ve_1 - q(\ve_1'+\ve_2') \right)
\eea
On the other hand, the second term $R_2$ is the contribution from all
the redistributions involving sites $j=2 \ldots N$, but not site $j=1$.
There are $(N-1)(N-2)/2$ such pairs of links, which all give the same
contribution to $R_2$. So $R_2$ can be written:
\bea
R_2 &=& \frac{N-2}{N} \, p(\ve_1) \, \int_0^{\infty} d\ve_2' \, p(\ve_2') \int_0^{\infty} d\ve_3' \, p(\ve_3') \times \\ \nonumber
&& \qquad \qquad \int_0^1 dq \, \psi(q) \int_0^{\infty} d\ve_2 \, \delta \left(\ve_2-q(\ve_2'+\ve_3')\right)
\eea
All the integrals in the r.h.s.~of the above equation give a 
contribution equal to unity, so that $R_2$ reduces to:
\be
R_2 = \left( 1-\frac{2}{N} \right) p(\ve_1)
\ee
Replacing the above results into Eq.~(\ref{eq-R1R2}), one finally 
obtains the following equation:
\bea
p(\varepsilon) &=& \int_0^{\infty} d\varepsilon_1 \, p(\varepsilon_1) \int_0^{\infty} d\varepsilon_2 \, p(\varepsilon_2) \times \\ \nonumber
&& \quad \int_0^1 dq \, \psi(q) \, \delta \left( \varepsilon - q(\varepsilon_1 + \varepsilon_2) \right)
\eea
which is precisely Eq.~(\ref{EM-MF}).

\end{document}